\documentclass{aastex}
\usepackage{spr-astr-addons}
\usepackage{url}
\usepackage{multirow}
\usepackage{enumerate}
\usepackage{booktabs}

\RequirePackage{color}

\begin{document}

\title{Investigation of the millisecond pulsar origins by their spin periods at the wavebands of radio, X-ray and $\gamma$-ray}
\slugcomment{Not to appear in Nonlearned J., 45.}
\shorttitle{Short article title}
\shortauthors{Wang et al.}

\author{De-Hua Wang}
\email{wangdh@gznu.edu.cn}
\affil{School of Physics and Electronic Science, Guizhou Normal University, Guiyang, 550001, China}

\and
\author{Cheng-Min Zhang}
\email{zhangcm@bao.ac.cn}
\altaffiltext{1}{National Astronomical Observatories, Chinese Academy of Sciences, Beijing, 100101, China}
\altaffiltext{2}{School of Physical Sciences, The University of Chinese Academy of Sciences, Beijing 101400, China}
\altaffiltext{3}{CAS Key Laboratory of FAST,  Chinese Academy of Sciences, Beijing 100101, China}
\altaffiltext{4}{Key Laboratory of Radio astronomy, Chinese Academy of Sciences, Beijing 100101, China}

\and
\author{Shuang-Qiang Wang}
\affil{Xinjiang Astronomical Observatory, Chinese Academy of Sciences, Urumqi, 830011, China}

\begin{abstract}
To track the formation and evolution links of the millisecond pulsars (MSPs) powered by accretion and rotation in the galactic field,
we investigate the spin period ($P$) and spin-down power ($\dot{E}$) distributions of the MSPs observed at the wavebands of radio, X-ray and $\gamma$-ray.
We find that all but one (119/120) of the $\gamma$-ray MSPs have been detected with the radio signals (radio+$\gamma$ MSPs), on the contrary, nearly half of the radio MSPs (118/237) have not been detected with $\gamma$-rays (radio-only MSPs). In addition, the radio+$\gamma$ MSPs are shown to be the relative faster and more energetic objects ($\langle P\rangle\sim3.28$\,ms  and   $\langle \dot{E}\rangle\sim4.5\times10^{34}\,{\rm erg\,s^{-1}}$) compared with the radio-only MSPs ($\langle P\rangle\sim4.70$\,ms and $\langle \dot{E}\rangle\sim1.0\times10^{34}\,{\rm erg\,s^{-1}}$), {\bf while the spin periods of these two MSP populations are   compatible with the log-normal distributions by the statistical tests}.
Most rotation-powered MSPs (RMSPs) with the radio eclipsing (31/34) exhibit the radio+$\gamma$ signals, which share the faster spin ($\langle P\rangle\sim2.78$\,ms) and larger spin-down power ($\langle \dot{E}\rangle\sim4.1\times10^{34}\,{\rm erg\,s^{-1}}$) distributions than the non-eclipsing ones ($\langle P\rangle\sim4.19$\,ms, $\langle \dot{E}\rangle\sim2.4\times10^{34}\,{\rm erg\,s^{-1}}$), implying the radio+$\gamma$ MSPs to be  younger than the radio-only MSPs.
It is noticed that the spin  distribution of the accretion-powered X-ray MSPs shows a clustering phenomenon around $\sim1.6-2.0$\,ms, which is not observed in RMSPs, hinting that the RMSPs may experience the multiple possible origins. Particularly, all the three super-fast spinning RMSPs with $P\sim1.4-1.6$\,ms exhibit the non-eclipsing, and we argue that they may be the distinctive sources formed by the accretion induced collapse (AIC) of white dwarfs.

\end{abstract}


\section{Introduction}\label{sec:intro}
\begin{table}
\caption{Accretion-powered X-ray MSPs and transitional MSPs in the galactic field.}
\begin{tabular}{@{}llccl@{}}
\noalign{\smallskip}\hline\hline\noalign{\smallskip}
\# & Source & $\nu_s$ & $P$ & Type$^\S$ \\
 & & (Hz) & (ms) & \\
\hline
\multicolumn{5}{l}{AMXP and NMXP in the galactic field} \\\relax
[1] & IGR J17602-6143 & 164 & 6.10 & A \\\relax
[2] & SWIFT J1756.9-2508 & 182 & 5.49 & A \\\relax
[3] & XTE J0929-314 & 185 & 5.41 & A \\\relax
[4] & XTE J1807.4-294 & 191 & 5.24 & A \\\relax
[5] & IGR J17511-3057 & 245 & 4.08 & A, N \\\relax
[6] & 4U 1916-05 & 270 & 3.70 & N \\\relax
[7] & IGR J17191-2821 & 294 & 3.40 & N \\\relax
[8] & XTE J1814-338 & 314 & 3.18 & A, N \\\relax
[9] & 4U 1702-429 & 330 & 3.03 & N \\\relax
[10] & 4U 1728-34 & 363 & 2.75 & N \\\relax
[11] & HETE J1900.1-2455 & 377 & 2.65 & A, N \\\relax
[12] & SAX J1808.4-3658 & 401 & 2.49 & A, N \\\relax
[13] & IGR J17498-2921 & 401 & 2.49 & A, N \\\relax
[14] & 4U 0614+09 & 415 & 2.41 & N \\\relax
[15] & XTE J1751-305 & 435 & 2.30 & A \\\relax
[16] & Swift J1749.4-2807 & 518 & 1.93 & A \\\relax
[17] & KS 1731-260 & 524 & 1.91 & N \\\relax
{\bf [18]} & {\bf IGR J17591-2342} & {\bf 527} & {\bf 1.90} & {\bf A} \\\relax
[19] & A 1744-361 & 530 & 1.89 & N \\\relax
{\bf [20]} & {\bf SAX J1810.8-2609} & {\bf 532} & {\bf 1.88} & {\bf N} \\\relax
[21] & Aql X-1 (1908+005) & 550 & 1.82 & A, N \\\relax
[22] & EXO 0748-676 & 552 & 1.81 & N \\\relax
[23] & MXB 1659-298 & 567 & 1.76 & N \\\relax
[24] & 4U 1636-53 & 581 & 1.72 & N \\\relax
[25] & MXB 1743-29 & 589 & 1.70 & N \\\relax
[26] & IGR J00291+5934 & 599 & 1.67 & A \\\relax
[27] & SAX J1750.8-2900 & 601 & 1.67 & N \\\relax
[28] & GS 1826-238 & 611 & 1.64 & N \\\relax
[29] & 4U 1608-52 & 619 & 1.62 & N \\
\hline
\multicolumn{5}{l}{tMSP in the galactic field} \\\relax
[1] & PSR J1023+0038 & 592 & 1.69 & \\\relax
[2] & PSR J1227-4853 & 593 & 1.69 & \\
\hline
\end{tabular}
\label{table::1}
\begin{tabular}{@{}l@{}}
$^\S$ A---accreting millisecond X-ray pulsar; \\N---nuclear-powered millisecond X-ray pulsar.
\end{tabular}
\end{table}

Based on the recycling interpretation for the formation of the millisecond pulsars (MSPs) \citep{Alpar82,Radhakrishnan82,Bhattacharya91}, the neutron star (NS) in a low-mass X-ray binary (LMXB) can accrete $\sim0.1-0.2\,{\rm M_\odot}$ \citep{Zhang11,Pan15} from its companion through the accretion disk during the $\sim0.1-10$\,Gyr   \citep{Tauris12}, then it is spun-up to a spin period of a few milliseconds, and probably also reduce its magnetic field strength to $\sim10^7-10^9$\,G \citep{Bhattacharya95,Zhang06,Zhang16}. The overall torque acting onto the NS during the spin up state depends on the disk structure, as well as the interaction between the NS magnetic field and the accretion plasma \citep{Ghosh79,Ghosh07,Kluzniak07}. After the X-ray accretion phase, the recycled   pulsar will change from the accretion-powered X-ray MSP into a rotation-powered MSP (RMSP) by emitting radio pulsation \citep{Lorimer08}.

Since 1990's, several observational evidence have been found to support the recycling scenario of MSP formation:
(1). The first evidence that constructs  the link between the accreting millisecond X-ray pulsar (AMXP, e.g., SAX J1808.4-3658) in LMXB and RMSP was detected by \citet{Wijnands98}, however,  it is suggested that some  AMXPs  may show the transition to the rotation-powered state during the X-ray quiescence \citep{Burderi06,Burderi09,DiSalvo08,Hartman08,Hartman09,Sanna17}.
(2). The transition between the accretion- and rotation-powered behaviors have been observed from IGR J18245-2452 \citep{Papitto13,Pallanca13,Ferrigno14,Linares14}, PSR J1023+0038 \citep{Archibald09,Stappers14,Patruno14}, and XSS J12270-4859 \citep{Bassa14,Papitto14,Roy14,Bogdanov14} (i.e., the transitional MPSs, or tMSPs, see \citealt{Papitto16}).
(3). The irregular radio eclipses are observed in some binary RMSPs (i.e., the eclipsing RMSPs including black widows and redbacks, see \citealt{Roberts13,Torres17}), which are explained as the absorptions by the lost matter ejected from the companions  \citep{Fruchter88,Kluzniak88}.

Until now, there have been  $\geq300$ RMSPs   detected (isolated and binary, see the ATNF pulsar catalogue   \citealt{Manchester05}), where the fastest one, i.e., PSR J1748-2446ad in the globular cluster, shows the spin frequency of 716\,Hz \citep{Hessels06}.  While,  $> 30$ {\bf accretion-powered X-ray MSPs}  \citep{Patruno17} have been detected, including the AMXPs with the spin signals observed from the accretion-powered coherent pulsations \citep{Wijnands98,Patruno12}, and the nuclear-powered millisecond X-ray pulsars (NMXPs) with the spin signals inferred from the thermonuclear burst oscillations \citep{Strohmayer96,Chakrabarty03,Strohmayer06,Watts12}. The details of the AMXPs {\bf , NMXPs} and tMSPs in the galactic field are shown in Table \ref{table::1}{\bf , including two new detected sources, i.e., IGR J17591-2342 \citep{Ferrigno18,Sanna18} and SAX J1810.8-2609 \citep{Bilous08}}. \citet{Papitto14} analyzed the spin distributions of AMXPs, NMSPs, eclipsing and non-eclipsing RMSPs, and find that NMXPs show the significantly faster spins than the most rotation-powered sources, while the eclipsing RMSPs show the faster spins than the non-eclipsing ones. Furthermore, \citet{Patruno17} indicated that there may exit two sub-populations in the spin frequency distributions of the AMXPs+NMXPs  with the mean values of $\approx300$\,Hz and $\approx575$\,Hz, respectively.
\begin{table*}
\footnotesize
\setlength{\tabcolsep}{1pt}
\caption{The MSP samples   in the galactic field.}
\begin{tabular}{@{}lccccl@{}}
\hline\hline\noalign{\smallskip}
Category & Count& Sub-count & Sub-sub-count & Fraction & Description \\
\hline
LMXBs & {\bf 29} & & & & All accretion-powered X-ray pulsars (AMXPs + NMXPs) \\
~~~AMXPs & & {\bf 14} & & 48\% & Accreting millisecond X-ray pulsars \\
~~~NMXPs & & {\bf 21} & & {\bf 72\%} & Nuclear-powered millisecond X-ray pulsars \\
\hline
RMSPs & 237 & & & & Rotation-powered (radio) millisecond pulsars \\
~~~eclipsing RMSPs & & 34 & & 14\% & RMSPs with irregularly eclipses in radio-pulsed emission  \\
~~~~~~radio+$\gamma$ MSPs & & & 31 & 13\% & Eclipsing RMSPs detected with both radio and $\gamma$-ray signals \\
~~~~~~radio-only MSPs & & & 3 & 1\% & Eclipsing RMSPs detected with radio but without $\gamma$-ray signals \\
~~~non-eclipsing RMSPs & & 203 & & 86\% & RMSPs whose radio-pulsed emission is not be eclipsed \\
~~~~~radio+$\gamma$ MSPs & & & 88 & 37\% & Non-eclipsing RMSP detected with both radio and $\gamma$-ray signals \\
~~~~~radio-only MSPs & & & 115 & 49\% & Non-eclipsing RMSP detected with radio but without $\gamma$-ray signals \\
\hline
radio-quiet $\gamma$-ray MSPs & 1 & & & 100\% & The MSP is detected with $\gamma$-ray signal but without radio signal\\
\hline
\end{tabular}
\label{table::2}
\end{table*}

In the times of $Fermi$ satellite, there are more than 100   MSPs   detected with $\gamma$-ray signals (from $\sim20$\,MeV to over 300\,GeV, see \citealt{Abdo13}), and these $\gamma$-ray MSPs tend to be the shorter-period, more energetic population than the canonical, non-recycled ones \citep{Ray12,Abdo13,Caraveo14,Grenier15}. Motivated by the analysis on the spin distributions of the MSPs by \citet{Papitto14} and \citet{Patruno17}, as well as the $\gamma$-ray observations for  RMSPs,  we try to compare the distributions of the spin period ($P$) and spin-down power ($\dot{E}$) of various MSPs at the different wavebands and powered by the accretion or spin-down, by which we may infer their origins, e.g.,  recycling process or accretion induced collapse (AIC) of white dwarf process.

The structure of the paper is organized as follows. In Section 2, we introduce the population of MSPs. Then in Section 3, we compare the $P$ and $\dot{E}$ distributions between the MSP samples with the different wavebands, e.g.,  radio, X-ray and $\gamma$-ray. Finally, we present the discussions and conclusions in Section 4.

\section{Population of millisecond pulsars}\label{sec:popu}
\begin{figure*}
\centering
\includegraphics[width=5.7cm]{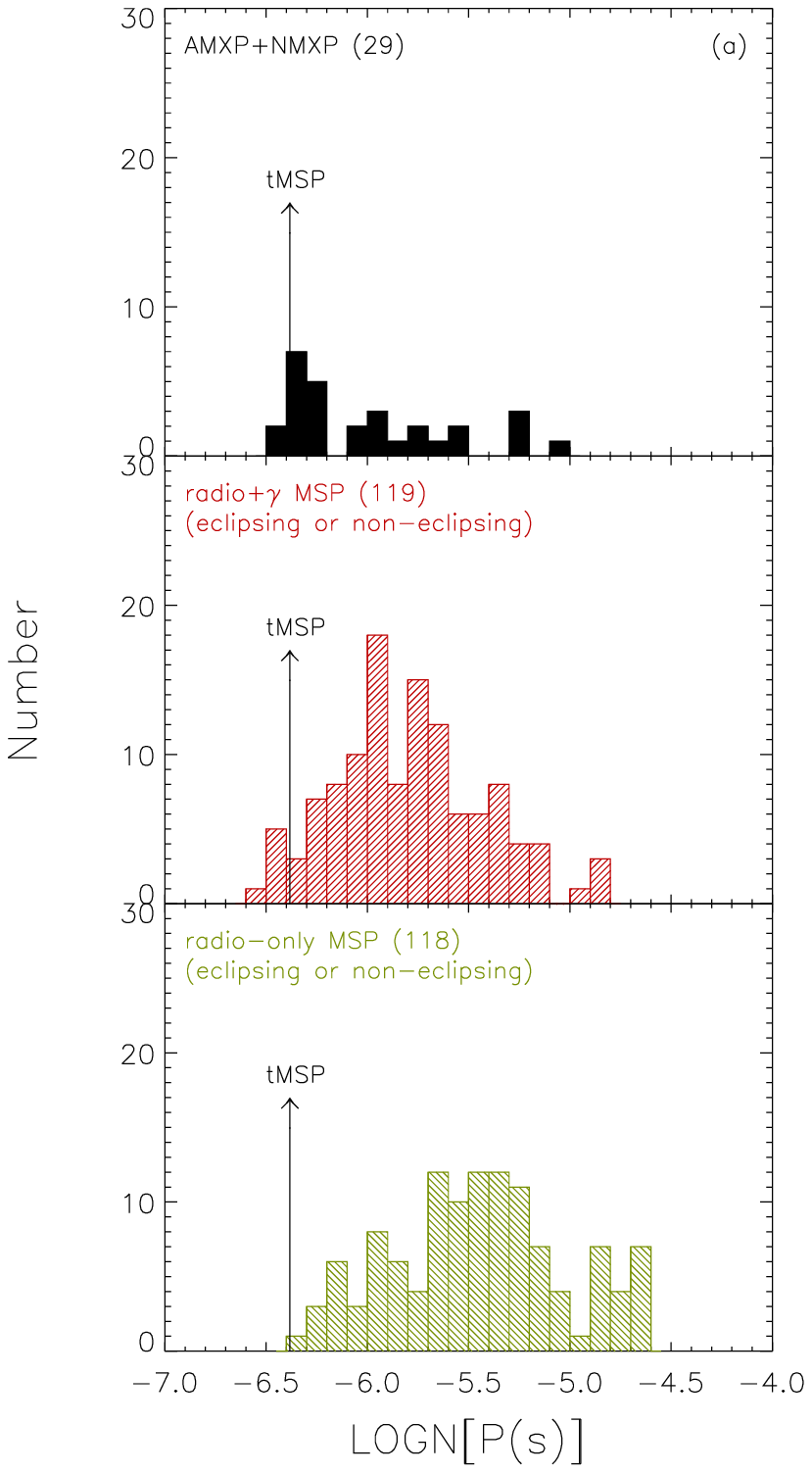}
\includegraphics[width=5.7cm]{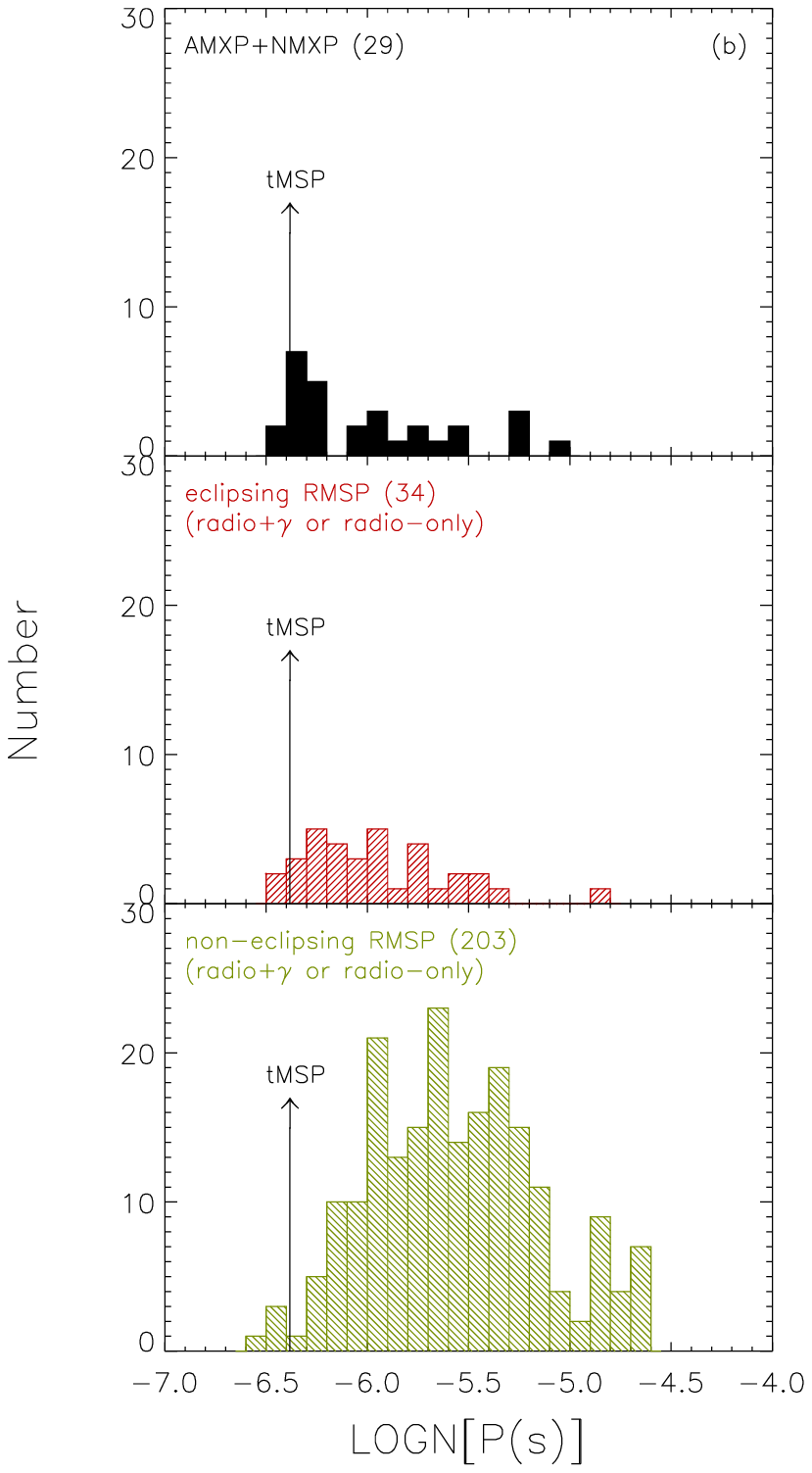}
\caption{Spin period distributions of the MSP samples {\bf (the horizontal axis stands for  the natural logarithm of $P$)}: (a) accretion-powered X-ray MSPs, radio+$\gamma$ and radio-only MSPs; (b) accretion-powered X-ray MSPs, eclipsing and non-eclipsing RMSPs. The spin periods of the tMSP PSR J1023+0038 ($P\sim1.69$\,ms) and PSR J1227-4853 ($P\sim1.69$\,ms) are also indicated in the figures.
}
\label{fig::1}
\end{figure*}
\begin{figure*}
\centering
\includegraphics[width=5.7cm]{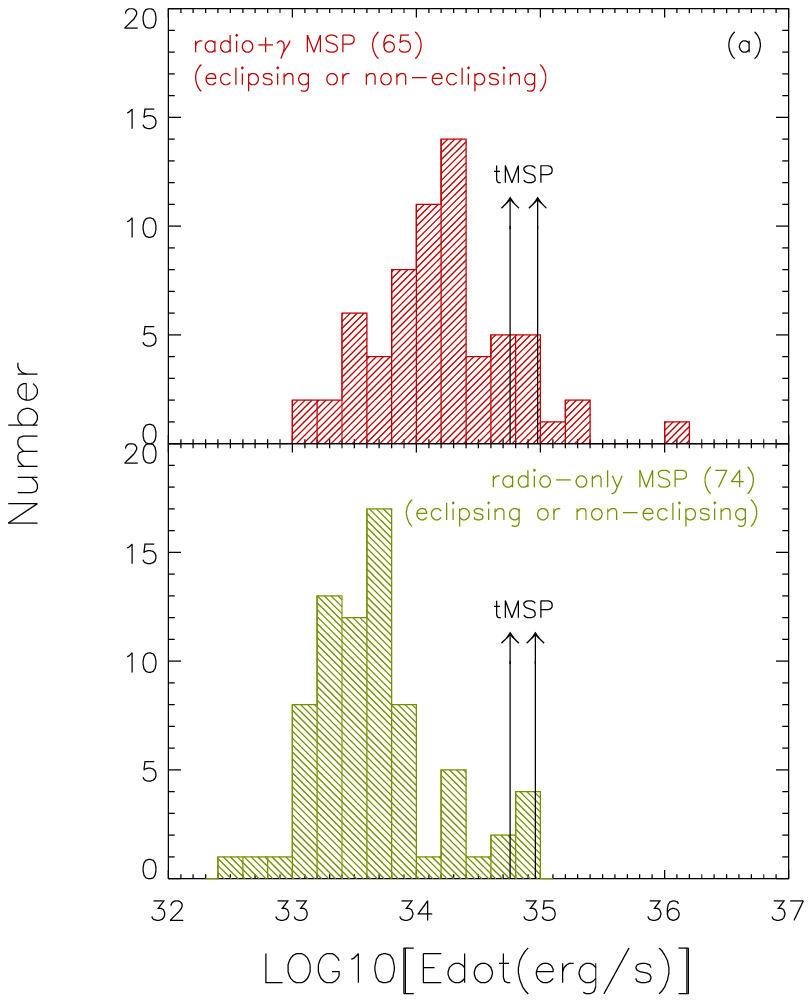}
\includegraphics[width=5.7cm]{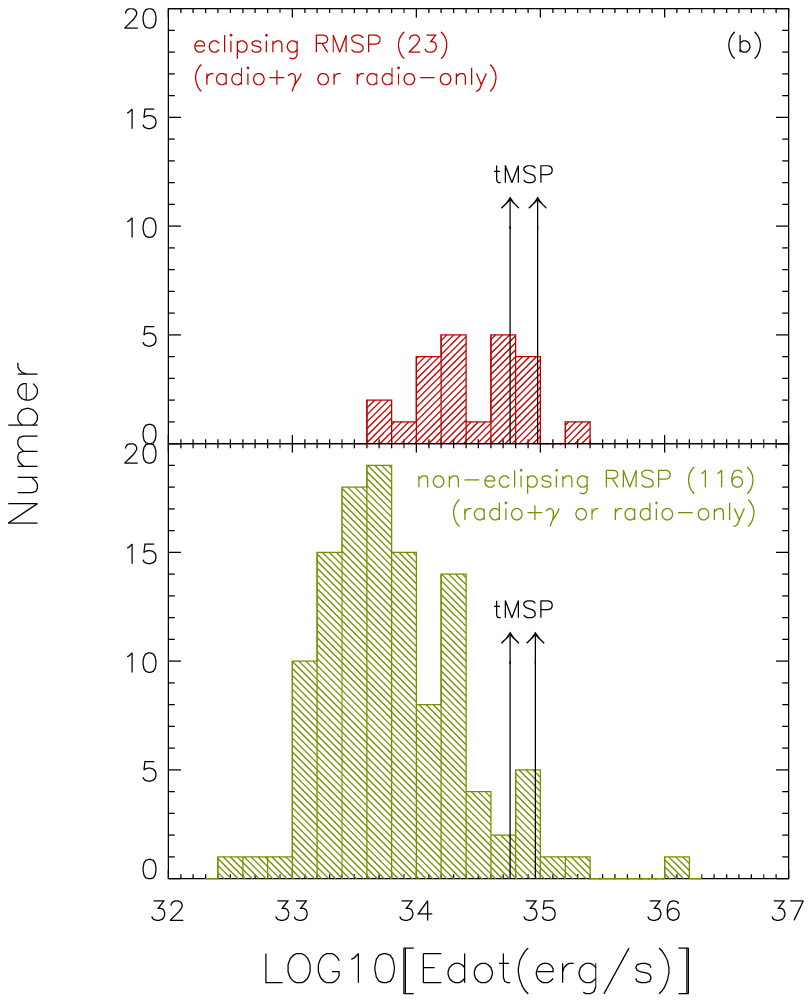}
\caption{Spin-down power distributions of the MSP samples: (a) radio+$\gamma$ and radio-only MSPs; (b) eclipsing and non-eclipsing RMSPs. The spin-down powers of the tMSP PSR J1023+0038
($\dot{E}\sim5.7\times10^{34}\,{\rm erg\,s^{-1}}$) and PSR J1227-4853 ($\dot{E}\sim9.1\times10^{34}\,{\rm erg\,s^{-1}}$) are also indicated in the figures.
}
\label{fig::2}
\end{figure*}

We follow the work by \citet{Papitto14} and \citet{Patruno17} to collect the MPS samples, including the accretion-powered X-ray MSPs (AMXPs+NMSPs), eclipsing and non-eclipsing RMSPs. However, this paper focuses on the comparisons of MSPs at  various radiation wavebands: AMXPs and NMXPs---X-ray, eclipsing and non-eclipsing RMSPs---radio or $\gamma$-ray. In addition, the selections of the MSP samples are also constrained by the following rules: (1). Only the MSPs in the galactic field are taken into account, while the ones in the globular cluster are excluded because  they may undergo the more complicated evolution processes.
(2). The MSP samples are selected with $P<10$\,ms, as it includes   most of the observed $\gamma$-ray MSPs. (3). Both the isolated and binary RMSPs are considered, since the progenitors of the isolated RMSPs must have gone through episodes of accretion (recycling) in their past history \citep{Patruno17}. (4) \citet{Papitto14} consider the "transitional MSPs" as the combination of AMXPs and eclipsing RMSPs, here instead, we follow \citet{Patruno17} and refer the transitional MSPs in the galactic field as the two systems shown in Table \ref{table::1}, for which there is actual evidence of a transition.

The details of the collected MSP samples are summarized in Table \ref{table::2}, where the accretion-powered X-ray MSPs are referred to \citet{Papitto14} and \citet{Patruno17}, the RMSPs are referred to the catalogs compiled by ATNF\footnote{http://www.atnf.csiro.au/research/pulsar/psrcat/} \citep{Manchester05} and D. R. Lorimer\footnote{http://astro.phys.wvu.edu/GalacticMSPs/GalacticMSPs.txt}, the $\gamma$-ray MSPs are referred to D. R. Lorimer$^3$ and "Public List of LAT-Detected Gamma-Ray Pulsars"\footnote{https://confluence.slac.stanford.edu/display/GLAMCOG/\\Public+List+of+LAT-Detected+Gamma-Ray+Pulsars}, and the eclipsing RMSPs are referred to "Millisecond Pulsar Catalogue"\footnote{https://apatruno.wordpress.com/about/millisecond-pulsar-catalogue/}. It is noticed that all but one of the $\gamma$-ray MSPs (119/120) have been detected with the radio signals, which are recorded as radio+$\gamma$ MSPs, on the contrary, nearly half of the radio MSPs (118/237) have not been detected with the $\gamma$-ray signals, which are recorded as radio-only MSPs. Besides, most eclipsing RMSPs (31/34) show radio+$\gamma$ signals, and two tMSPs in the galactic field (both are eclipsing RMSPs), i.e., PSR J1023+0038 and PSR J1227-4853, have shown the transition from the X-ray emission in the accretion-powered stage to the radio emission in the rotation-powered stage, where  PSR J1227-4853 has also been detected with $\gamma$-ray pulsation in the rotation-powered stage.  There is only one MSP (PSR J1744-7619) that has been detected with $\gamma$-ray signal but without radio signal ($<30\,{\rm \mu Jy}$, see \citealt{Abdo13}), i.e., the radio-quiet $\gamma$-ray MSP, which shows the spin period of $\sim4.7$\,ms \citep{Clark18}.

\begin{table}
\caption{Spin period statistics of the MSP samples.}
\footnotesize
\setlength{\tabcolsep}{4pt}
\begin{tabular}{@{}lccccc@{}}
\hline\hline\noalign{\smallskip}
Category & Count & Range & $\langle P\rangle^a$ & $\tilde{P}^b$ & $\sigma_P^c$ \\
\noalign{\smallskip}
\cline{3-6}
 & & \multicolumn{4}{c}{(ms)} \\
\hline
LMXBs & {\bf 29} & $1.62-6.10$ & {\bf 2.75} & {\bf 2.30} & {\bf 1.33} \\
\hline
radio+$\gamma$ MSPs  & 119 & $1.41-8.12$ & 3.28 & 2.96 & 1.34 \\
radio-only MSPs & 118 & $1.69-9.90$ & 4.70 & 4.20 & 2.05 \\
\hline
eclipsing RMSPs & 34 & $1.61-7.61$ & 2.78 & 2.48 & 1.20 \\
non-eclipsing RMSPs & 203 & $1.41-9.90$ & 4.19 & 3.68 & 1.88 \\
\hline
\end{tabular}
\begin{tabular}{@{}l@{}}
$^a$ $\langle P\rangle$: mean of $P$; \\
$^b$ $\tilde{P}$: median of $P$; \\
$^c$ $\sigma_P$: standard deviation of $P$. \\
\end{tabular}
\label{table::3}
\end{table}
\begin{table}
\caption{$K-S$ test$^a$ results of the spin period {\bf and spin-down power} distributions.}
\footnotesize
\setlength{\tabcolsep}{4pt}
\begin{tabular}{@{}lccccl@{}}
\hline\hline\noalign{\smallskip}
Category & Count & $K-S$ & Reject $H_0$ \\
 & & ($p$-value) & \\
\hline\noalign{\smallskip}
\multicolumn{4}{l}{{\bf Spin Persiod ($P$)}} \\
LMXBs & {\bf 29} & \multirow{2}*{${\bf 2.02\times10^{-3}}$} & \multirow{2}*{yes} \\
radio+$\gamma$ MSPs & 119 & & \\
\hline
LMXBs & {\bf 29} & \multirow{2}*{${\bf 1.87\times10^{-6}}$} & \multirow{2}*{yes} \\
radio-only MSPs & 118 & & \\
\hline
radio+$\gamma$ MSPs  & 119 & \multirow{2}*{$5.34\times10^{-9}$} & \multirow{2}*{yes} \\
radio-only MSPs  & 118 & & \\
\hline\hline\noalign{\smallskip}
\multicolumn{4}{l}{{\bf Spin-down Power ($\dot{E}$)}} \\
{\bf radio+$\gamma$ MSPs} & {\bf 65} & \multirow{2}*{${\bf 1.58\times10^{-9}}$} & \multirow{2}*{{\bf yes}} \\
{\bf radio-only MSPs} & {\bf 74} & & \\
\hline
{\bf eclipsing RMSPs} & {\bf 23}  & \multirow{2}*{${\bf 5.35\times10^{-6}}$} & \multirow{2}*{{\bf yes}} \\
{\bf non-eclipsing RMSPs} & {\bf 116}  & & \\
\hline
\end{tabular}
\begin{tabular}{@{}l@{}}
\begin{minipage}{80mm}
$^a$ {\bf $H_0$ is the null hypothesis that the two groups of data are from the same continuous distribution,
with the confidence level parameter $\alpha=0.05$}. \\
\end{minipage}
\end{tabular}
\label{table::4}
\end{table}
\begin{table}
\caption{{\bf $S-W$ test$^a$ results of the spin period distribution.}}
\footnotesize
\setlength{\tabcolsep}{4pt}
\begin{tabular}{@{}lccc@{}}
\hline\hline\noalign{\smallskip}
{\bf Category} & {\bf Count} & {\bf $S-W$} & {\bf Reject $H_0$} \\
 & & {\bf ($p$-value)} \\
\hline
\noalign{\smallskip}\multicolumn{4}{l}{{\bf Normality}} \\
{\bf radio+$\gamma$ MSPs}  & {\bf 119} & ${\bf 7.63\times10^{-8}}$ & {\bf Yes} \\
\hline
{\bf radio-only MSPs} & {\bf 118} & ${\bf 1.76\times10^{-6}}$ & {\bf Yes} \\
\hline\hline
\noalign{\smallskip}\multicolumn{4}{l}{{\bf Log-Normality}} \\
{\bf radio+$\gamma$ MSPs}  & {\bf 119} & ${\bf 2.51\times10^{-1}}$ & {\bf No} \\
\hline
{\bf radio-only MSPs} & {\bf 118} & ${\bf 9.87\times10^{-2}}$ & {\bf No} \\
\hline
\end{tabular}
\begin{tabular}{@{}l@{}}
\begin{minipage}{80mm}
$^a$ {\bf $H_0$ is the null hypothesis that the data follows a normal or a log-normal distribution,
with  the confidence level parameter $\alpha=0.05$.} \\
\end{minipage}
\end{tabular}
\label{table::add}
\end{table}

\section{Comparison of the $P$ and $\dot{E}$ distributions}\label{sec:comp}
We collect $P$ and $\dot{E}$ data of the MSP samples, and then compare their distributions among  the various MSP categories classified by the wavebands, e.g., radio, X-ray and $\gamma$-ray.

\subsection{The distribution of $P$}
\begin{table*}
\caption{Spin-down power statistics of the MSP samples.}
\begin{tabular}{@{}lccccc@{}}
\hline\hline\noalign{\smallskip}
Category & Count & Range & $\langle \dot{E}\rangle^a$ & $\tilde{\dot{E}}^b$ & $\sigma_{\dot{E}}^c$ \\
\noalign{\smallskip}\cline{3-6}\noalign{\smallskip}
 & & \multicolumn{4}{c}{($\rm erg\,s^{-1}$)} \\
\hline
radio+$\gamma$ MSPs  & 65 & $1.4\times10^{33}-1.1\times10^{36}$ & $4.5\times10^{34}$ & $1.5\times10^{34}$ & $1.4\times10^{35}$ \\
radio-only MSPs & 74 & $2.8\times10^{32}-9.8\times10^{34}$ & $1.0\times10^{34}$ & $4.4\times10^{33}$ & $1.9\times10^{34}$ \\
\hline
eclipsing RMSPs & 23 & $5.3\times10^{33}-1.6\times10^{35}$ & $4.1\times10^{34}$ & $2.5\times10^{34}$ & $3.8\times10^{34}$ \\
non-eclipsing RMSPs & 116 & $2.8\times10^{32}-1.1\times10^{36}$ & $2.4\times10^{34}$ & $5.3\times10^{33}$ & $1.0\times10^{35}$ \\
\hline
\end{tabular}\\
\begin{tabular}{@{}l@{}}
$^a$ $\langle \dot{E}\rangle$: mean of $\dot{E}$; \\
$^b$ $\tilde{\dot{E}}$: median of $\dot{E}$; \\
$^c$ $\sigma_{\dot{E}}$: standard deviation of $\dot{E}$. \\
\end{tabular}
\label{table::5}
\end{table*}

The spin period statistics, including the range, mean ($\langle P\rangle$), median ($\tilde{P}$) and standard deviation ($\sigma_P$), of the various MSP categories are summarized in Table \ref{table::3}, and Figure \ref{fig::1} shows the corresponding histograms.
It can be seen that the radio+$\gamma$ MSPs (total 119) show the $P$ distribution ($\langle P\rangle\sim3.28$\,ms and $\tilde{P}\sim2.96$\,ms) intermediate between the radio-only MSPs (total 118, $\langle P\rangle\sim4.70$\,ms and $\tilde{P}\sim4.20$\,ms) and the accretion-powered X-ray MSPs (total {\bf 29}, {\bf $\langle P\rangle\sim2.75$\,ms} and {\bf $\tilde{P}\sim2.30$\,ms}). The $Kolmogorov-Smirnov$ ($K-S$) test shows that the spin periods of these three types of MSPs come from the different continuous distribution at the 95 percent confidence level, as shown in Table \ref{table::4}.

{\bf We obtain the similar conclusions to those of  \citet{Patruno17} that the accretion-powered X-ray MSPs show a clustering phenomenon in the $P$ distribution around $\sim1.6-2.0$\,ms (the corresponding spin frequency is $\sim500-600$\,Hz, see Figure \ref{fig::1}). In addition, } it should be also noticed from Table \ref{table::3} and Figure \ref{fig::1}(a) that the radio-only MSPs and the accretion-powered X-ray MSPs show the similar minimal spin periods of $P\sim1.6$\,ms, however, three radio+$\gamma$ MSPs show the even faster spins: PSR J0952-0607 ($P\sim1.41$\,ms, see \citealt{Bassa17}), PSR J1803+1358 ($P\sim1.52$\,ms, see the catalog from D. R. Lorimer\footnote{http://astro.phys.wvu.edu/GalacticMSPs/GalacticMSPs.txt}) and PSR J1939+2134 (B1937+21, $P\sim1.56$\,ms, see \citealt{Backer82}).
{\bf Some analysis argued that the spin periods of the RMSPs may be from a population with a log-normal distribution \citep{Lorimer15}, but not with a normal distribution \citep{Tauris12,Papitto14}. Here we take the $Shapiro-Wilk$ ($S-W$) test to check the $P$ distributions of the radio+$\gamma$ and radio-only MSPs, and find that both  populations show the spin periods  to be  incompatible with a normal distribution at the 95 percent confidence level, but compatible with a log-normal distribution with $(\mu\pm\sigma)_{\rm radio+\gamma}\sim(-5.79\pm0.37){\rm \,log_e(s)}$ and $(\mu\pm\sigma)_{\rm radio-only}\sim(-5.45\pm0.43){\rm \,log_e(s)}$, respectively (see Table \ref{table::add}).
}

Furthermore, the similar results to \citet{Papitto14} and \citet{Patruno17} can be obtained from Table \ref{table::3} and Figure \ref{fig::1}(b) that the eclipsing RMSPs (total 34) show the $P$ distribution ($\langle P\rangle\sim2.78$\,ms and $\tilde{P}\sim2.48$\,ms) faster than the non-eclipsing RMSPs (total 203, $\langle P\rangle\sim4.19$\,ms and $\tilde{P}\sim3.68$\,ms).  It should be also noticed that all  the three fastest radio+$\gamma$ MSPs, i.e., PSR J0952-0607, PSR J1803+1358 and PSR J1939+2134 (B1937+21) have not been reported with the observed radio eclipsing phenomena.

It is convenient to take the two tMSPs in the galactic field, i.e., PSR J1023+0038 and PSR J1227-4853, as a reference to check the $P$ and $\dot{E}$ distributions  of  the MSP samples. Figure \ref{fig::1} shows that their spin periods are same (both $P\sim1.69$\,ms, see also Table \ref{table::1}), which are faster than the  other MSP samples, but slower than  the three fastest radio+$\gamma$ MSPs.

\subsection{The distribution of $\dot{E}$}

The spin-down power statistics, including the range, mean ($\langle \dot{E}\rangle$), median ($\tilde{\dot{E}}$) and standard deviation ($\sigma_{\dot{E}}$), of the various MSP categories are summarized in Table \ref{table::5}, and Figure \ref{fig::2} shows the corresponding histograms.

It can be seen from Table \ref{table::5}, Figure \ref{fig::2}(a) and Figure \ref{fig::2}(b) that the radio+$\gamma$ MSPs (total 65) show the $\dot{E}$ distribution ($\langle \dot{E}\rangle\sim4.5\times10^{34}\,{\rm erg\,s^{-1}}$ and $\tilde{\dot{E}}\sim1.5\times10^{34}\,{\rm erg\,s^{-1}}$) larger than the radio-only MSPs (total 74, $\langle \dot{E}\rangle\sim1.0\times10^{34}\,{\rm erg\,s^{-1}}$ and $\tilde{\dot{E}}\sim4.4\times10^{33}\,{\rm erg\,s^{-1}}$), while the eclipsing RMSPs (total 23) show the $\dot{E}$ distribution ($\langle \dot{E}\rangle\sim4.1\times10^{34}\,{\rm erg\,s^{-1}}$ and $\tilde{\dot{E}}\sim2.5\times10^{34}\,{\rm erg\,s^{-1}}$) larger than the non-eclipsing RMSPs (total 116, $\langle \dot{E}\rangle\sim2.4\times10^{34}\,{\rm erg\,s^{-1}}$ and $\tilde{\dot{E}}\sim5.3\times10^{33}\,{\rm erg\,s^{-1}}$). {\bf The $K-S$ test indicates that the $\dot{E}$ of the radio+$\gamma$ and radio-only MSPs come from the different continuous distributions at the 95 percent confidence level, while the $\dot{E}$ of the eclipsing and non-eclipsing RMSPs also come from the different continuous distributions, as shown in Table \ref{table::4}.}

The two tMSPs, i.e., PSR J1023+0038 and PSR J1227-4853, show the $\dot{E}$ of $\sim5.7\times10^{34}\,{\rm erg\,s^{-1}}$ and $\sim9.1\times10^{34}\,{\rm erg\,s^{-1}}$ respectively, which are larger than those of most other MSP samples (see Figure \ref{fig::2}).  In addition, as expected, the fast rotator  PSR J1939+2134 (B1937+21) with $P\sim1.56$\,ms shows the large $\dot{E}$ of $\sim1.1\times10^{36}\,{\rm erg\,s^{-1}}$.

\section{Discussions and Conclusions}\label{sec:dis}

We compare the $P$ and $\dot{E}$ distributions among  various types of MSPs in the   galactic field, including the accretion-powered X-ray MSPs (AMXPs+NMSPs), eclipsing and non-eclipsing RMSPs, and focus on their radiative wavebands. The details of the discussions and conclusions are summarized as below:

\begin{itemize}
\item The count of the radio+$\gamma$ MSPs (119) collected in the paper is comparable to that of the radio-only MSPs (118, see Table \ref{table::2}), and the radio+$\gamma$ MSPs tend to be the shorter-period ($\langle P\rangle\sim3.28$\,ms), more energetic ($\langle \dot{E}\rangle\sim4.5\times10^{34}\,{\rm erg\,s^{-1}}$) population than the radio-only MSPs ($\langle P\rangle\sim4.70$\,ms and $\langle \dot{E}\rangle\sim1.0\times10^{34}\,{\rm erg\,s^{-1}}$, see Table \ref{table::3}, Table \ref{table::5}, Figure \ref{fig::1}(a) and Figure \ref{fig::2}(a)). \citet{Arons96} suggests that due to some threshold voltage, the $\gamma$-ray luminosity $L_\gamma$ of the pulsar may relate to $\dot{E}$ as $L_\gamma\propto\dot{E}^{1/2}$, which is basically supported by the observations from $Fermi$ \citep{Abdo13}. For a magnetic dipole model of the pulsar, combining the relations of $\dot{E}\sim(32{\rm \pi}^4/3c^3)(B^2R^6/P^4)\propto P^{-4}$ and $L_\gamma\propto\dot{E}^{1/2}$ will derive $L_\gamma\propto P^{-2}$, which may explain why the MSPs with the faster $P$ or larger $\dot{E}$ are more likely to emit $\gamma$-rays. In fact, all the $\gamma$-ray MSP samples in the paper show $P<10$\,ms and $\dot{E}>10^{33}\,{\rm erg\,s^{-1}}$. It is also noticed that most eclipsing RMSPs (31/34, see Table \ref{table::2}) show radio+$\gamma$ signals, which share the faster $P$ ($\langle P\rangle\sim2.78$\,ms) and larger $\dot{E}$ ($\langle \dot{E}\rangle\sim4.1\times10^{34}$) distributions than the non-eclipsing ones ($\langle P\rangle\sim4.19$\,ms and $\langle \dot{E}\rangle\sim2.4\times10^{34}$, see Table \ref{table::3}, Table \ref{table::5}, Figure \ref{fig::1}(b) and Figure \ref{fig::2}(b)). Sine it is suggested that the eclipsing RMSPs may link to their accreting progenitors \citep{Kluzniak88}, so we suspect that the radio+$\gamma$ MSPs may be younger than the radio-only MSPs.

\item {\bf The $K-S$ tests indicate that the radio+$\gamma$ and radio-only MSPs share the different $P$ and $\dot{E}$ distributions (see Table \ref{table::4}). In addition, the $S-W$ tests verify that the $P$ distributions of these two MSP populations are both compatible with being log-normal (see Table \ref{table::add}). It should be noticed that the above conclusions  depend on the sample selection introduced in section 2, however, we still suggest that there should be a physical difference between the radio+$\gamma$ and radio-only MSPs. So far, it has been  neither clear whether there is an evolutional relation between the two MSP populations, nor what physical process dominates the difference between them, which need further observational and theoretical analysis.
    }

\item It can be seen from Figure \ref{fig::1} that many accretion-powered X-ray MSPs {\bf (14/29)} show the $P$ distribution clustering around $\sim1.6-2.0$\,ms, as similar to the result by \citet{Patruno17}. This phenomenon can be explained by the spinning  limit due to some effect which acts as a "brake" on the NS spins, such as the gravitational wave radiation \citep{Bildsten98,Andersson99,Chakrabarty03,Chakrabarty08,Haskell11,Guo16,Patruno17}, the NS magnetic field \citep{Patruno12} and the transient accretion \citep{Bhattacharya17,DAngelo17}. However, the similar clustering phenomenon is not observed in the $P$ distribution of RMSPs (see Figure \ref{fig::1}). Moreover, three non-eclipsing RMSPs emitting radio+$\gamma$ signals, i.e., PSR J0952-0607, PSR J1803+1358 and PSR J1939+2134 (B1937+21), share the even faster $P$ of $\sim1.4-1.6$\,ms (see Figure \ref{fig::1}). It is not clear why the accretion-powered X-ray MSPs with the fast spin of $P<1.6$\,ms, as corresponding to these three RMSPs, have not been detected (the fastest spin of the accretion-powered X-ray MSP is about 1.62\,ms, see \citealt{Hartman03}). We rather suggest that the clustering distribution around $P\sim1.6-2.0$\,ms shown in the accretion-powered X-ray MSPs may be due to the selective effect of the limited samples.

\item As an example, we take the two tMSPs in the galactic field, i.e., PSR J1023+0038 and PSR J1227-4853, as a reference to check the above results of the $P$ and $\dot{E}$ distributions. Firstly, the two sources are in the end phase of the accretion-powered stage with $P\sim1.69$\,ms, which is in the clustering area around $\sim1.6-2.0$\,ms shown in $P$ distribution of the accretion-powered X-ray MSPs (see Figure \ref{fig::1}), implying that this spin range may relate to the transitional process between the accretion- and rotation-powered stage. Moreover, the two tMSPs are the new born RMSPs with observed irregularly radio eclipses, supporting that the radio eclipsing phenomenon may link to their accreting progenitors.

\end{itemize}

The observations from multi-wavebands are critical for  understanding  the MSP evolution between the accretion- and rotation-powered stages, and the conclusions in this paper may provide some clues for the further investigations.
In addition, the $K-S$ test shows that RMSPs and the accretion-power X-ray MSPs share the different $P$ distributions, implying that RMSPs is unlikely to have evolved from a single coherent progenitor population, and this fact has been noticed by \citet{Kiziltan09}. Furthermore, we find that all the three observed super-fast spinning  RMSPs  with $P\sim1.4-1.6$\,ms  exhibit  the non-eclipsing, so we argue that
they may be the distinctive MSPs experiencing other origins, such as the accretion induced collapse of white dwarfs
\citep{Bhattacharya91,Nomoto95,Taani12,Kiziltan13}.

\acknowledgments
This work is supported by the National Natural Science Foundation of China (Grant No. 11703003 and No. U1731238), the Science and Technology Foundation of
Guizhou Province (Grant No. [2017]5726), NAOC-Y834081V01.



\end{document}